\documentstyle[aps,12pt]{revtex}
\newcommand{\refindent}{\par\penalty-100\noindent
               \hangindent 20pt\hangafter=1\null}
\begin{document}
\draft
\title{Black Holes\footnote{To appear in the American Physical Society
Centenary issue of Reviews of Modern Physics, March 1999.}}
\author{Gary T. Horowitz}
\address{Physics Department, University of California, Santa Barbara, CA 93106}
\author{Saul A. Teukolsky}
\address{Departments of Physics and Astronomy, Cornell University,
Ithaca, NY 14853}
\maketitle
\bigskip
\begin{abstract}
\baselineskip 12pt
Black holes are among  the most intriguing objects in modern physics. Their
influence ranges from powering quasars and other active galactic nuclei,
to providing key insights into quantum gravity. We
review the observational evidence for black holes, and briefly discuss
some of their properties. We also describe some recent developments
involving cosmic censorship and the statistical origin of black hole 
entropy.
\end{abstract}

\baselineskip 16pt

\section{Introduction}

Black holes are predicted by general relativity to be formed
whenever sufficient mass is compressed into a small enough volume.
In Newtonian language, the escape velocity from the surface becomes
greater than the speed of light, so that nothing can escape.
In general relativity, a black hole is defined as a region of
spacetime that cannot communicate with the external universe. The
boundary of this region is called the surface of the black hole,
or the event horizon.

It appears impossible to compress matter on earth sufficiently  to
form a black hole. But in nature, gravity itself can compress matter
if there is not enough pressure to resist the inward attractive force.
When a massive star reaches the endpoint of its thermonuclear burning
phase, nuclear reactions no longer supply thermal pressure, and
gravitational collapse will proceed all the way to a black
hole. By contrast,
the collapse of a less massive star halts at high density when the
core is transformed entirely into nuclear matter. The envelope of the
star is blown off in a gigantic supernova explosion, leaving the core
behind as a nascent neutron star.

The ``modern" history of the black hole begins with the classic
paper of Oppenheimer and Snyder (1939). They calculated the collapse
of a homogeneous sphere of pressureless gas in general relativity.
They found that the sphere eventually becomes cut off from all
communication with the rest of the universe. Ultimately the
matter is crushed to infinite density at the center.
Most previous discussions of the exterior gravitational field
of a spherical mass had not understood that the apparent
singularity in the solution at the Schwarzschild radius was
merely a coordinate artifact. Einstein himself claimed
that one needn't worry about the ``Schwarzschild singularity"
since no material body could ever be compressed to such a radius
(Einstein 1939). His error was that he considered only
bodies in equilibrium. Even the usually sober Landau
had been bothered by the prospect of
continued gravitational collapse implied by the existence of
a maximum stable mass for neutron stars and white dwarfs.
To circumvent this, he believed at one time that
``\ldots all stars heavier than $1.5M_\odot$ certainly possess
regions in which the laws of quantum mechanics \ldots are violated"
(Landau 1932).

Despite the work of Oppenheimer and Snyder, black holes were
generally ignored until the late 1950's, when Wheeler and his
collaborators began a serious investigation of the problem
of gravitational collapse (Harrison {\it et al.} 1965).
It was Wheeler (1968) who coined the name ``black hole."
The discovery of quasars, pulsars, and compact X-ray
sources in the 1960's finally gave observational
impetus to the subject, and ushered in the ``golden age" of
black hole research.

Black holes are  now believed to exist with a variety of masses. 
A current estimate for the dividing line 
between progenitor stars that produce
neutron stars and those that produce black holes is 
 around $25M_\odot$.
The resulting black holes are expected to have masses
in the range $3$ -- $60\,M_\odot$.
As discussed
below, there is also good astrophysical evidence for supermassive
black holes, with masses of order $10^6$ -- $10^9\,M_\odot$.
There are a number of scenarios that could produce such large black holes:
the gravitational collapse of individual supermassive gas clouds;
the growth of a seed black hole capturing stars and gas from
a dense star cluster at the center of a galaxy; or
the merger of smaller black holes produced by collapse.
There have also been
speculations that black holes with a very wide range of masses might
have been produced from density fluctuations in the early universe, but
so far there is no convincing evidence for the existence of such primordial
black holes.

This article provides just an overview of
the astrophysical evidence for black holes, and discusses
some recent theoretical developments in black hole research.
For a more complete discussion of the basic properties of black holes, see the
books by Misner, Thorne, and Wheeler (1973), Shapiro and Teukolsky (1983), or
Wald (1984).

\section{Observational Evidence for Black Holes}

\subsection{The Maximum Mass of Neutron Stars}

Neutron stars can exist happily in equilibrium for small enough masses.
But beyond a certain critical mass, the inward pull of gravity overwhelms
the balancing pressure force---the star is unstable to collapsing
to a black hole. This provides one of the key observational signatures
of a black hole astronomically: Look for a system containing a dark,
compact object. If you can determine that the mass of the object is
greater than the maximum allowed mass of a neutron star, then it
must be a black hole.

The value of the maximum neutron star mass is uncertain theoretically
because we don't understand nuclear physics well enough to calculate
it reliably (see, e.g., Baym 1995).
Current conventional nuclear equations of state predict
a maximum mass around $2M_\odot$ (see, e.g. discussion and references
in Cook, Shapiro, and Teukolsky 1994 or Baym 1995). (For some ``unconventional"
possibilities, see Brown and Bethe 1994; Bahcall, Lynn, and Selipsky
1990; Miller, Shahbaz, and Nolan 1997).

Because of these uncertainties, astrophysicists generally rely on
a calculation that assumes we understand nuclear physics up to
some density $\rho_0$, and then varies the pressure-density relation over
all possibilities beyond this point to maximize the resulting mass
(Rhoades and Ruffini 1974).
This procedure yields an upper limit to the maximum
mass of
\begin{equation}\label{rhoades}
M_{\rm max}\simeq 3.2M_\odot\left({4.6\times
 10^{14}\,{\rm g\,cm^{-3}}\over \rho_0} \right)^{1/2}.
\end{equation}
Kalogera and Baym (1996) have redone the Rhoades-Ruffini calculation
with more modern physics and obtained essentially the same numbers:
a coefficient of $2.9M_\odot$ for a preferred matching density of
$5.4\times 10^{14}\,\rm g\, cm^{-3}$.
Rotation increases the amount of matter
that can be supported against collapse, but only by about 25\% even
for stars rotating near breakup speed (see, e.g., Cook, Shapiro,
and Teukolsky 1994). The Rhoades-Ruffini calculation assumes
a causality condition, that the speed of sound is less than
the speed of light: $dP/d\rho \leq c^2$. Abandoning this
assumption increases the coefficient in eq.\ (\ref{rhoades})
from 3.2 to 5.2 (Hartle and Sabbadini 1977 and references
therein). But it is not clear that this can be done without
the material of the star becoming spontaneously unstable
(Bludman and Ruderman 1970; but see Hartle 1978).
In summary, circumventing these mass limits
would require us to accept some unconventional physics---much more
unconventional than black holes!

\subsection{Observational Signatures of Black Holes}\label{signature}

A black hole is the most compact configuration of matter possible for
a given mass. The size of a black hole of mass $M$ is given by the
Schwarzschild radius, the radius of the event horizon:
\begin{equation}\label{rs}
R_S={2GM\over c^2}=3\,{\rm km} \left({M\over M_\odot}\right).
\end{equation}
One way of verifying the compactness of a candidate black hole is by
measuring the speed of matter in orbit around it, which is expected to
approach $c$ near the horizon.  This test is feasible
since accretion flows of orbiting gas
are common around gravitating objects in astrophysics.
In a few objects, direct evidence for high
orbital speeds is obtained by measuring the Doppler broadening of
spectral lines from the accreting gas. More often, black hole candidates
exhibit gas outflows, or jets, with relativistic speeds.
Another indication of
compactness comes from observations of strong X-ray emission from the
accreting gas, which imply high temperatures $>10^9$ K.
Such temperatures are easily achieved by accretion onto
a black hole or a neutron star, both of which have sufficiently
deep potential wells.

When the radiation (typically X-rays) from a compact object varies on
a characteristic time scale $t$, without contrived conditions the size of
the object must be less than $ct$.
If this size limit is
comparable to $R_S$ (determined from an independent mass estimate)
then the object is a
potential black hole. For solar mass black holes, this implies
looking for variability on the scale of less than a millisecond.

The demonstration of compactness alone, however, is not sufficient to
identify a black hole; a neutron star, with a radius of about $3R_S$,
is only slightly larger than a black hole of the same mass. Clear
evidence that $M > M_{\rm max}$ is needed in addition to compactness.

Any gravitating object has a maximum luminosity, the Eddington limit,
given by
\begin{equation}\label{edd}
L_{\rm Edd}\simeq 10^{38}{\,\rm erg\,s^{-1}}\left({M\over M_\odot}\right)
\end{equation}
(see, e.g., Shapiro and Teukolsky 1983).
Above this
luminosity, the outward force due to escaping radiation on the accreting gas
overwhelms the attractive force due to gravity, and accretion is no
longer possible. Thus the observed luminosity sets a lower limit on
the mass of the accreting object, which can often suggest the presence
of a black hole.

\subsection{Supermassive Black Holes in Galactic Nuclei}

Quasars emit immense amounts of radiation, up to
$\sim10^{46}~{\rm erg\,s^{-1}}$, from very small volumes.
They are members of a wider class of objects, active
galactic nuclei (AGN), all of which generally radiate intensely.

Nearly all AGN emit substantial fractions of their radiation in
X-rays, and some emit the bulk of their radiation in even more energetic
$\gamma$-rays.  Rapid variability of the flux has been observed in
some AGN.  Many AGN also have relativistic jets.  These are all
signatures of a compact relativistic object.
If the observed radiation is powered by
accretion, as generally assumed, then the Eddington limit
(\ref{edd}) implies
masses in the range $10^6-10^{10}M_\odot$.  This is
well above the maximum mass of a neutron star, and so AGN are considered
secure black hole candidates.
Menou, Quataert, and Narayan (1997) give a summary of the current best
supermassive black hole candidates 
at the centers of nearby galaxies.

Direct evidence for the existence of a central relativistic potential well
has come from the recent detection of broad iron fluorescence lines in
X-rays in a few AGN.  The line broadening
can be interpreted as a combination of Doppler broadening and
gravitational redshift. A spectacular example is the galaxy
MCG-6-30-15, where a very broad emission line has been observed.
The data can be interpreted as suggesting that the central
mass is a rapidly rotating black hole, but this is still
tentative. (See Menou, Quataert, and Narayan 1997 for discussion
and references for this source and many others. See Rees 1998 for
a general discussion of astrophysical evidence for black holes.)

\subsection{Black Holes in X-Ray Binaries}

In an X-ray binary one of the stars is compact and accretes gas from the outer
layers of its companion.
Because of angular momentum conservation in the rotating system, gas
cannot flow directly onto the compact object. Instead, it spirals towards
the compact object and heats up because of viscous dissipation, producing
X-rays.
In many cases, the compact star is
known to be a neutron star, but there are also a number of excellent
black hole candidates.
 
The mass of the X-ray emitting star $M_X$ can be constrained by
observations of the spectral lines of the secondary star.  The
Doppler shifts of these lines give an estimate of the radial velocity $v_r$
of the secondary as it orbits the X-ray star.  Combining $v_r$
with the orbital period $P$ of the binary and using
Kepler's third law yields the ``mass function'' of the compact object,
\begin{equation}\label{mass}
f(M_X)\equiv\frac{M_X \sin ^3 i}{(1+q)^2}={Pv_r^3\over 2\pi G}.
\end{equation}
(see, e.g., Shapiro and Teukolsky 1983).
The mass function does not give $M_X$ directly
because of its dependence on the unknown inclination $i$ of the
binary orbit and the ratio $q$ of the two masses.  However, it is a
firm {\it lower} limit on $M_X$.  Therefore, mass functions above $3$
$M_{\odot}$ suggest the presence of black holes.  Additional
observational data---absence or presence of eclipses, for instance,
or information on the nature of the secondary star---can help to
constrain $i$ or $q$, so that a likely value of $M_X$ can often be
determined. The best stellar mass black hole candidates currently
known are summarized in
Menou, Quataert, and Narayan (1997).

The first black hole candidate discovered in this way was Cyg X--1.
Although its mass function is not very large, there are good observations
that set limits on $i$ and $q$ and suggest that $M_X$ is definitely
greater than 3 -- 4$M_\odot$, with the likely value being 7 -- 20$M_\odot$.
Even stronger evidence is provided by other X-ray binaries
for which $f(M_X)>3M_\odot$. Without any further astrophysical assumptions,
one can be pretty sure that these objects are not  neutron stars. Currently
the most compelling black hole candidate is V404 Cyg,
with a mass function of $6M_\odot$.

Many of these sources show the key observational signatures of
black holes described in Section \ref{signature}.
Some display rapid variability in their X-ray emission. Many
occasionally reach high luminosities, implying masses greater
than that of a neutron star via the Eddington limit (\ref{edd}).
A few exhibit relativistic jets.

\subsection{Conclusive Evidence for Black Holes}

All the methods for finding black holes described above
are indirect. They essentially say that there is a lot
of mass in a small volume.
Direct proof that a candidate object is a black hole requires a
demonstration that the object has the spacetime geometry predicted
by Einstein's theory. For example, we would like to have evidence
for an event horizon, the one feature
that is unique to a black hole.

One possible approach is via accretion theory (see Menou, Quataert, and
Narayan 1997 for a review).
Two kinds of accretion are important for flow onto compact objects.
The first is accretion in a thin disk. The accreting gas quickly
radiates whatever energy is released through viscous dissipation.
The gas stays relatively cool and so the disk remains thin, each
gas element orbiting the central mass at the Keplerian velocity.
Unlike the Newtonian case, the gravitational field of a compact
mass in general relativity has a last stable circular orbit.
The inner edge of the disk extends up to this radius. Observations
such as those of the iron fluorescence lines described above
provide information on the radius of the inner edge of the
accretion disk. Since the radius of the last stable circular
orbit depends on the spin of central mass, we may be able to
measure the spin of black holes in this way.

Thin disks have oscillatory modes whose details depend on general
relativity. Quasi-periodic oscillations have been detected in
several X-ray binaries, and can be used to probe the spacetime
geometry (``diskoseismology"; see Rees 1998 for a review and
references). In addition, if the disk is tilted
with respect to the spin axis of the central mass, it will precess
because of frame dragging (Lense-Thirring effect). This produces
a periodic modulation of the X-ray luminosity, which may already
have been seen in a few cases.

The second important kind of accretion is the advection-dominated
accretion flow (ADAF). Here the accreting gas advects most of the
energy released by viscosity to the center. The gas becomes
relatively hot and quasi-spherical. The spectrum is quite different
from that of a thin disk. ADAFs appear to be present in both galactic
nuclei and in X-ray binaries when the accretion rate is relatively
low. In an ADAF, what happens to the energy advected to the center
depends on the nature of the central object. If it is a black hole,
the energy simply disappears behind the event horizon. If
it is a neutron star or any object with a surface, the energy
is reradiated from the surface and will dominate the spectrum.
For those black hole candidates that seem to be accreting
in ADAFs, the evidence is that they lack surfaces.
While not yet conclusive because of the modeling uncertainties,
this is the most direct evidence yet that black holes with
event horizons are present in nature.

Is there any hope of a clean observation of a black hole geometry,
without the complications of dirty astrophysics? The best hope
is from the observation of gravitational waves from
black hole collisions (see the article by Weiss in this volume).
Laser interferometers now under
construction, such as LIGO, VIRGO, and GEO (see, e.g., Abramovici
{\it et al.} 1992; Thorne 1994) will be sensitive to black hole-black hole
and black hole-neutron star collisions with black hole masses
up to a few tens of solar masses. The predicted event rate for
such collisions is highly uncertain: estimates range from about one per year for
the initial LIGO detector and thousands
per year for the upgraded LIGO (Siggurdson and Hernquist 1993; Lipunov,
Postnov, and Prokhorov 1997; Bethe and Brown 1998), to essentially
zero (Zwart and Yungelson 1998). If nature is kind and we do detect
such events, the wave form encodes a great deal of information
about the spacetime geometry. The part of the wave form from the
highly nonlinear merger phase is currently being calculated
with large scale supercomputer simulations (see, e.g., Finn 1997), and
it is expected that comparison of such calculations with observations
should yield not only the masses and spins of the colliding
objects, but also a check that the wave form is consistent
with general relativity. The final part of the wave form is
a ``ring down", like a damped harmonic oscillator. It has been
calculated by perturbation theory, and should provide another
strong test.

There is also good reason to believe that when two galaxies
each containing supermassive black holes merge, the black holes
will spiral together and coalesce. The frequency of the gravitational
waves emitted is too low to be detectable on earth, where
the waves would be swamped by seismic noise.
However, such events should be readily detectable
by a laser interferometer in space, such as the proposed LISA
detector (see, e.g., Bender {\it et al.} 1996)

\section{Black Hole Uniqueness}

The solution of Einstein's equations that describes a spherical black
hole was discovered by Karl Schwarzschild only a few months after
Einstein published the final form of general relativity:
\begin{equation}\label{schw}
ds^2=-\left(1-{2M\over r}\right)dt^2+\left(1-{2M\over r}\right)^{-1}
dr^2+r^2\,(d\theta^2+\sin^2\theta\,d\phi^2).
\end{equation}
(Here, and for the remainder of our discussion, we use units with $c=G=1$.)
This metric turns out to be the only spherically symmetric solution
in the absence of matter. In general relativity, as in Newtonian
gravity,
the vacuum gravitational field outside any spherically symmetric
object is the same as that of a point mass. The event horizon occurs
at $r=2M$ (cf.\ eq.\ \ref{rs}). Although the metric components are singular
there, they can be made regular by a simple change of coordinates.
In contrast, the singularity at $r=0$ is real. An observer falling
into a Schwarzschild black hole will be ripped apart by infinite
tidal forces at $r=0$.

One might expect that solutions of Einstein's equations describing
realistic black holes that form in nature and settle down to
equilibrium would be very complicated. After all, a black hole can
be formed from collapse of all kinds of matter configurations, with
arbitrary multipole distributions, magnetic fields, distributions
of angular momentum, and so on. For most situations, after the black hole
has settled down,  
it can be described by a solution of Einstein's
vacuum field equations. Remarkably,  one can show that the only stationary
solution of this equation that is asymptotically flat and has a regular
event horizon is a generalization of (\ref{schw}) known as
the Kerr metric. This solution has only two parameters:
the mass $M$  and angular momentum $J$. 
All other information about the precursor
state of the system is radiated away during the collapse.
Astrophysical black holes are not expected to have a large electric
charge since 
free charges are rapidly neutralized by plasma in an astrophysical
environment. Nevertheless, there is an analog of this uniqueness theorem
for charged black holes: all stationary solutions of the Einstein-Maxwell
equations that are asymptotically flat and have a  regular
event horizon are known, and depend only $M$, $J$ and the charge $Q$.

The simplicity of the final black hole state is summarized by
Wheeler's aphorism, ``A black hole has no hair." This is supported not
only by the above uniqueness theorems, but also by results showing
that if one couples general relativity to simple matter fields e.g. free
scalar fields, there are no new stationary black hole solutions.
However, it has recently been
shown that if more complicated matter is considered, new black hole solutions
can be found. Examples include Einstein-Yang-Mills black holes,
black holes inside magnetic monopoles, and charged black holes coupled
to a scalar ``dilaton".  Even these new black holes are characterized by
only a few parameters, so the spirit of Wheeler's aphorism is maintained.
(For a recent review and references, 
see Bekenstein 1997.)

\section{Cosmic Censorship}

In the late 1960's, a series of powerful results were established in 
general relativity which 
show that under generic conditions, gravitational collapse produces
infinite gravitational fields, i.e.,  infinite spacetime curvature
(see, e.g., Hawking and Ellis 1973).
However, these ``singularity theorems" do not guarantee the existence
of an event horizon. It is known that uniform density,
spherically symmetric gravitational
collapse produces a black hole (the Oppenheimer-Snyder
solution), and small perturbations 
do not change this. It is  conceivable, however, that 
highly nonspherical collapse or, e.g., the collision of two black holes
could produce singularities that are not hidden behind event horizons.
These regions of infinite curvature 
would be visible to distant observers and hence are
called ``naked" singularities. Penrose (1969) proposed that
naked singularities could not form in realistic situations, a hypothesis that
has become known as cosmic censorship. If this is violated, general relativity
could break down outside black holes, and would not be
sufficient to predict the future evolution. On the positive side, this would
open up the
possibility of direct observations of quantum gravitational effects.
Establishing whether cosmic censorship
holds is perhaps the most important open question in classical
general relativity today.

Despite almost thirty years of effort, we are still far from a general
proof of cosmic censorship. (For a recent review and references, see Wald 1997.)
This seems to require analysis of the late
time evolution of Einstein's equation in the strong field regime.
The much simpler problem of determining the
global evolution of relatively weak (but still nonlinear) gravitational
waves was only achieved in the late 1980's, and hailed as a technical
tour-de-force. 
In light of this, progress has been made by studying simpler
systems, trying to find counterexamples, and by numerical simulations.
The simpler systems are usually
general relativity with one or two symmetries imposed. For example, cosmic 
censorship has been established for a class of solutions with two commuting
symmetries.
One class of potential counterexamples consists of time symmetric initial data
containing a minimal surface $S$. Assuming cosmic censorship, one can show
that the area of this minimal surface must be related to the total mass $M$
by $A(S) \le 16\pi M^2$. Unsuccessful attempts were made to find
initial data that
violate this inequality. Recently, a general proof of this inequality has
been found, showing that no counterexamples of this type exist.
Numerical simulations of nonspherical collapse have found some indication
that cosmic censorship may be violated in certain situations (Shapiro
and Teukolsky 1991),
and suggest
that any theorem might need careful specification of what is meant by
``generic" initial data. 

Perhaps the most effort, and most interesting results, have come from studying
spherically symmetric collapse. It was shown in the early 1970's that
naked singularities could form in inhomogeneous dust collapse, but it was
quickly realized that these ``shell crossing" or ``shell focusing" 
singularities also occurred in the absence of gravity and just reflected
an unrealistic model of matter. It was believed at the time that any
description of matter that did not produce singularities in flat spacetime
would not produce naked singularities when coupled to gravity. This has
recently been shown to be false. Consider spherically symmetric scalar
fields coupled to gravity. If the initial amplitude is small, the waves
will scatter and disperse to infinity. If the initial amplitude is large,
the waves will collapse to form a black hole. As one continuously varies
the amplitude, there is a critical value that divides these two outcomes.
It has been shown that at this critical value, the evolution produces a naked
singularity. This is not believed to be a serious counterexample to cosmic
censorship since it is not generic. But it  again indicates that a true
formulation of cosmic censorship is rather subtle.

Studies of spherical scalar field collapse near the critical amplitude
${\cal A}_0$ have yielded a surprising result. The mass of the resulting
black hole, for ${\cal A}>{\cal A}_0$, is
 \begin{equation}\label{scale}
 M_{BH} \sim |{\cal A} -{\cal A}_0|^\gamma
\end{equation}
where $\gamma $ is a universal exponent that is independent of the initial
wave profile.
Gravitational collapse of other matter fields, or axisymmetric
gravitational waves, exhibit similar
behavior (with a different exponent). Furthermore, the solution with
${\cal A}={\cal A}_0$, exhibits a type of scale invariance. These properties are similar to 
critical phenomena in condensed matter systems. They are not yet fully
understood, but may turn out to be related to thermodynamic properties of
black holes, which we discuss next. For recent reviews of critical
phenomena in gravitational collapse, see Gundlach (1998) and Choptuik (1998).

\section{Quantum Black Holes}

For an equilibrium black hole, one can define a
quantity called the surface gravity $\kappa$ which can be thought of as
the force that must be 
exerted on a rope at infinity to hold a unit mass stationary near the
horizon of a black hole. 
During the early 1970's, it was shown that black holes have the following
properties: 

\begin{description}
\item[(0)] The surface gravity is constant over the horizon, even
for rotating black holes which are not spherically symmetric. 
\item[(1)] If one
throws a small amount of mass into a stationary black hole characterized by
$M, Q, J$, it will settle down to a new  stationary black hole. The change
in these three quantities satisfies
\begin{equation} \label{thermo}
\delta M = {\kappa \delta A\over 8\pi} + \Omega \delta J
\end{equation}
where $A$ is the area of the event horizon and $\Omega$ is the angular
velocity of the horizon. 
\item[(2)] The  area of a black hole cannot decrease during physical processes.
\end{description}

It was immediately noticed that there was a close similarity between
these ``laws of black hole mechanics" and the usual laws of thermodynamics,
with $\kappa$ proportional to the temperature and $A$ proportional to the
entropy. However it was originally thought that this could only be an analogy,
since if a black hole really had a nonzero temperature, it would have to
radiate and everyone knew  that nothing could escape from a black hole.
 This view changed completely
when Hawking (1975) showed that if matter is treated  quantum mechanically,
black holes do radiate. This showed that black holes are indeed
thermodynamic objects with a temperature and entropy given by
\begin{equation}
 T_{bh} = {\hbar \kappa \over 2\pi}, \qquad
S_{bh} = {A\over 4\hbar}. 
\end{equation}
This turns out to be an enormous
entropy, much larger than the entropy of a corresponding amount of ordinary
matter. For a review of black hole thermodynamics, see Wald (1998).

In all other
contexts, we know that thermodynamics is the result of averaging over a large
number of different microscopic configurations with the same macroscopic
properties.  So it is natural to ask, what are the microstates of a black hole
that are responsible for its thermodynamic properties?
This  question has recently been answered
in both of the dominant approaches to quantum
gravity today: string theory and canonical quantization of general relativity.
We will focus on the situation in string theory, since this is further 
developed.  (String theory 
is discussed in more detail in the article by  Schwarz and
Seiberg in this volume.) Briefly, it is based
on the idea that elementary particles are not pointlike,
but actually different excitations of a one-dimensional extended
object---the string. Strings interact by a simple splitting and joining
interaction that turns out to reproduce the standard interactions of 
elementary particles. The strength of the interactions is governed by
a string coupling constant $g$. A crucial ingredient in string theory is that it
is supersymmetric. In any supersymmetric theory, the mass and charge
satisfy an inequality of the form $M\ge cQ$ for some constant $c$. 
States that saturate this bound are called BPS states and have
the special property that their mass does not receive any quantum corrections.

Now consider all BPS states in string theory with a given large charge $Q$.
At weak string coupling $g$, these states are easy to describe and count. Now
imagine increasing the string coupling. This increases the force of gravity,
and causes these states to become black holes. Charged black holes 
 also satisfy the inequality $M\ge cQ$, and when 
equality holds, the black holes are called extremal. So the BPS states
all become extremal black holes. But there is only one black hole for
a given mass and charge, so the BPS states all become identical black holes.
This is the origin of the thermodynamic properties of black holes. When
one compares the number of BPS states $N$ to the area of the event horizon,
one finds that in the limit of large charge 
\begin{equation}
N=e^{S_{bh}}
\end{equation}
in precise agreement with black hole thermodynamics. This agreement has
been shown to hold for near extremal black holes as well, where the mass
is slightly larger than $cQ$.

Extremal black holes have zero Hawking temperature and hence do not radiate.
But near extremal black holes do radiate approximately thermal radiation
at low temperature. Similarly, the interactions between near BPS states
in string theory produce  radiation. Remarkably, it turns out that the
radiation predicted in string theory agrees precisely with that coming from
black holes. This includes deviations from the black body spectrum, which
arise from two very different sources in the two cases. In the black hole
case, the deviations occur because the radiation has to 
propagate through the curved spacetime around the black hole. This gives
rise to an effective potential that results in a frequency dependent
``grey body factor" in the radiation spectrum. The string calculation at
weak coupling is done in flat spacetime so there are no curvature corrections.
Nevertheless, there are deviations from a purely thermal spectrum 
because there are separate left- and right-moving degrees of
freedom  along the string. Remarkably, the resulting spectra agree.
Progress has also been made in understanding the entropy of black holes
far from extremality. In both string theory and a canonical quantization
of general relativity there are calculations of the entropy of neutral
black holes up to an undetermined numerical
coefficient. For reviews of these developments in string theory see
Horowitz (1998) or Maldacena (1996). For the canonical quantization
results, see Ashtekar et. al. (1998).

\section{Conclusion}

Black holes connect to a wide variety of fields of physics. They are
invoked to explain high-energy phenomena in astrophysics, they are
the subject of analytic and numerical inquiry in classical general
relativity, and they may provide key insights into quantum gravity.
We also seem to be on the verge of verifying that
these objects actually exist in nature with the spacetime properties
given by Einstein's theory.  Finding absolutely
incontrovertible evidence for a black hole would be the
capstone of one of the most remarkable discoveries in the history of
science.

\acknowledgements

We thank Ramesh Narayan for helpful discussions. This work was supported in
part by NSF Grants PHY95-07065 at Santa Barbara and PHY 94-08378 at
Cornell University.

\section*{References}

\refindent
Abramovici, A., W. E. Althouse, R. W. P. Drever, Y. Gursel,
S. Kawamura, F. J. Raab, D. Shoemaker, L. Sievers, R. E. Spero,
K. S. Thorne, R. E. Vogt, R. Weiss, S. E. Whitcomb, and M. E.
Zucker, 1992, Science {\bf 256}, 325.
\refindent
Ashtekar, A.,  J. Baez, A. Corichi, K. Krasnov, 1998, Phys. Rev. Lett.
{\bf 80}, 904.
\refindent
Baym, G., 1995, Nucl. Phys. A {\bf 590}, 233c.
\refindent
Bahcall, S., B. W. Lynn, and S. B. Selipsky, 1990, Astrophys. J. {\bf 362},
251.
\refindent
Bekenstein, J. D., 1997, in {\it Second International A. D. Sakharov
Conference on Physics}, ed. I. M. Dremin and A. Semikhatov (World
Scientific, Singapore); expanded version in gr-qc/9605059.
\refindent
Bender, P., {\it et al.}, 1996, LISA Pre-Phase A Report, Max-Planck-Institut
fur Quantenoptik, Report MPQ 208, Garching, Germany. Available at
http://www.mpq.mpg.de/mpq-reports.html.
\refindent
Bethe, H. A., and G. E. Brown, 1998, astro-ph/9802084.
\refindent
Bludman, S. A., and M. A. Ruderman, 1970, Phys. Rev. D {\bf 1}, 3243.
\refindent
Brown, G. E., and H. A. Bethe, 1994, Astrophys. J. {\bf 423}, 659.
\refindent
Choptuik, M., 1998, gr-qc/9803075.
\refindent
Cook, G. B., S. L. Shapiro, and S. A. Teukolsky, 1994, Astrophys. J. {\bf
424}, 823.
\refindent
Einstein, A., 1939, Ann. Math. {\bf 40}, 922.
\refindent
Finn, L. S., 1997, in {\it Proceedings of the 14th
International Conference on General Relativity and Gravitation}, ed.
M. Francaviglia, G. Longhi, L. Lusanna, and E. Sorace (World Scientific,
Singapore), p.~147. Also gr-qc/9603004.
\refindent
Gundlach, C., 1998, to appear in Adv. Theor. Math. Phys.; also
gr-qc/9712084.
\refindent
Harrison, B. K., K. S. Thorne, M. Wakano, and J. A. Wheeler, 1965, {\it
Gravitation Theory and Gravitational Collapse} (University of Chicago
Press, Chicago).
\refindent
Hartle, J. B., 1978, Phys. Rep. {\bf 46}, 201.
\refindent
Hartle, J. B., and A. G. Sabbadini, 1977, Astrophys. J. {\bf 213}, 831.
\refindent
Hawking, S., 1975, Commun. Math. Phys. {\bf 43}, 199.
\refindent
Hawking, S. W., and G. F. R. Ellis, 1973, {\it The Large Scale
Structure of Space-time} (Cambridge University Press, Cambridge).
\refindent
Horowitz, G. T., 1998, in {\it Black Holes and Relativistic Stars},
 ed. R. M. Wald
(University of Chicago Press, Chicago), p.~241.
\refindent
Kalogera, V., and G. Baym, 1996, Astrophys. J. Lett. {\bf 470}, L61.
\refindent
Landau, L. D., 1932, Phys. Z. Sowjetunion {\bf 1}, 285.
\refindent
Lipunov, V. M., K. A. Postnov, and M. E. Prokhorov, 1997, New Astron. {\bf 2},
43.
\refindent
Maldacena, J. M., hep-th/9607235.
\refindent
Menou, K., E. Quataert, and R. Narayan, 1997, to appear in
{\it Proceedings of the Eighth Marcel Grossman Meeting on
General Relativity}; also astro-ph/9712015.
\refindent
Miller, J. C., T. Shahbaz, and L. A. Nolan, 1997, submitted to
Mon. Not. R. Astron. Soc.; also astro-ph/9708065.
\refindent
Misner, C. W., K. S. Thorne, and J. A. Wheeler, 1973, {\it
Gravitation} (Freeman, San Francisco).
\refindent
Oppenheimer, J. R., and H. Snyder, 1939, Phys. Rev. {\bf 56}, 455.
\refindent
Penrose, R., 1969, Riv. Nuovo Cim. {\bf 1}, 252.
\refindent
Rees, M. J., 1998, in {\it Black Holes and Relativistic Stars}, ed.
R. M. Wald (University of Chicago Press, Chicago), p.~79.
\refindent
Rhoades, C. E., and R. Ruffini, 1974, Phys. Rev. Lett. {\bf 32}, 324.
\refindent
Shapiro, S. L., and S. A. Teukolsky, 1983, {\it Black Holes, White
Dwarfs, and Neutron Stars: The Physics of Compact Objects} (Wiley,
New York).
\refindent
Shapiro, S. L., and S. A. Teukolsky, 1991, Phys. Rev. Lett. {\bf 66},
994.
\refindent
Siggurdson, S., and L. Hernquist, 1993, Nature {\bf 364}. 423.
\refindent
Thorne, K. S., 1994, in {\it Relativistic Cosmology, Proceedings
of the 8th Nishinomiya-Yukawa Memorial Symposium}, ed. M. Sasaki
(Universal Academy Press, Tokyo), p.~67.
\refindent
Wald, R. M. 1984, {\it General Relativity}, (University of Chicago Press,
Chicago).
\refindent
Wald, R. M., 1997, gr-qc/9710068.
\refindent
Wald, R. M., 1998, {\it Black Holes and Relativistic Stars}, ed. R. M. Wald
(University of Chicago Press, Chicago), p.~155.
\refindent
Wheeler, J. A., 1968, American Scientist {\bf  56}, 1.
\refindent
Zwart, S. F. P.,  and L. R. Yungelson, 1998, Astron. Astrophys. in press;
also astro-ph/9710347.

\end{document}